\documentclass[final,1p,times]{elsarticle}
\makeatletter
\def\ps@pprintTitle{%
  \let\@oddhead\@empty
  \let\@evenhead\@empty
  \def\@oddfoot{}%
  \let\@evenfoot\@oddfoot
}
\makeatother

\usepackage{amssymb}
\usepackage{amsmath}
\usepackage[table,xcdraw]{xcolor}
\usepackage{graphicx}
\usepackage{caption}
\usepackage{subcaption}
\usepackage{hyperref}
\usepackage{textcomp}
\usepackage{microtype}

\journal{Digital Signal Processing}
\biboptions{sort&compress}
\begin{document}

\begin{frontmatter}



\title{CPSNet: Physics-Inspired Label-Free Deep Unfolding for Lung Ultrasound B-Line Detection}

\author[label1,label2]{Tianqi Yang\corref{cor1}\fnref{fn1}} 

\cortext[cor1]{Corresponding author.}

\author[label3]{Oktay~Karakuş}
\author[label2]{Nantheera Anantrasirichai}
\author[label4]{Marco Allinovi}
\author[label2]{Alin~Achim}

\affiliation[label1]{organization={UCL Hawkes Institute},
            addressline={University College London}, 
            city={London},
            postcode={WC1V 6LJ}, 
            country={UK}}

\affiliation[label2]{organization={Visual Information Laboratory},
            addressline={University of Bristol}, 
            city={Bristol},
            postcode={BS8 1UB}, 
            country={UK}}
\affiliation[label3]{organization={School of Computer Science and Informatics},
            addressline={Cardiff University}, 
            city={Cardiff},
            postcode={CF24 4AG}, 
            country={UK}}
\affiliation[label4]{organization={Nephrology, Dialysis and Transplantation},
            addressline={Careggi University Hospital}, 
            city={Florence},
            country={Italy}}
            
\fntext[fn1]{The author is currently with the Hawkes Institute. This work was conducted while the author was a PhD student at the University of Bristol.}

\begin{graphicalabstract}
\includegraphics[width=\linewidth]{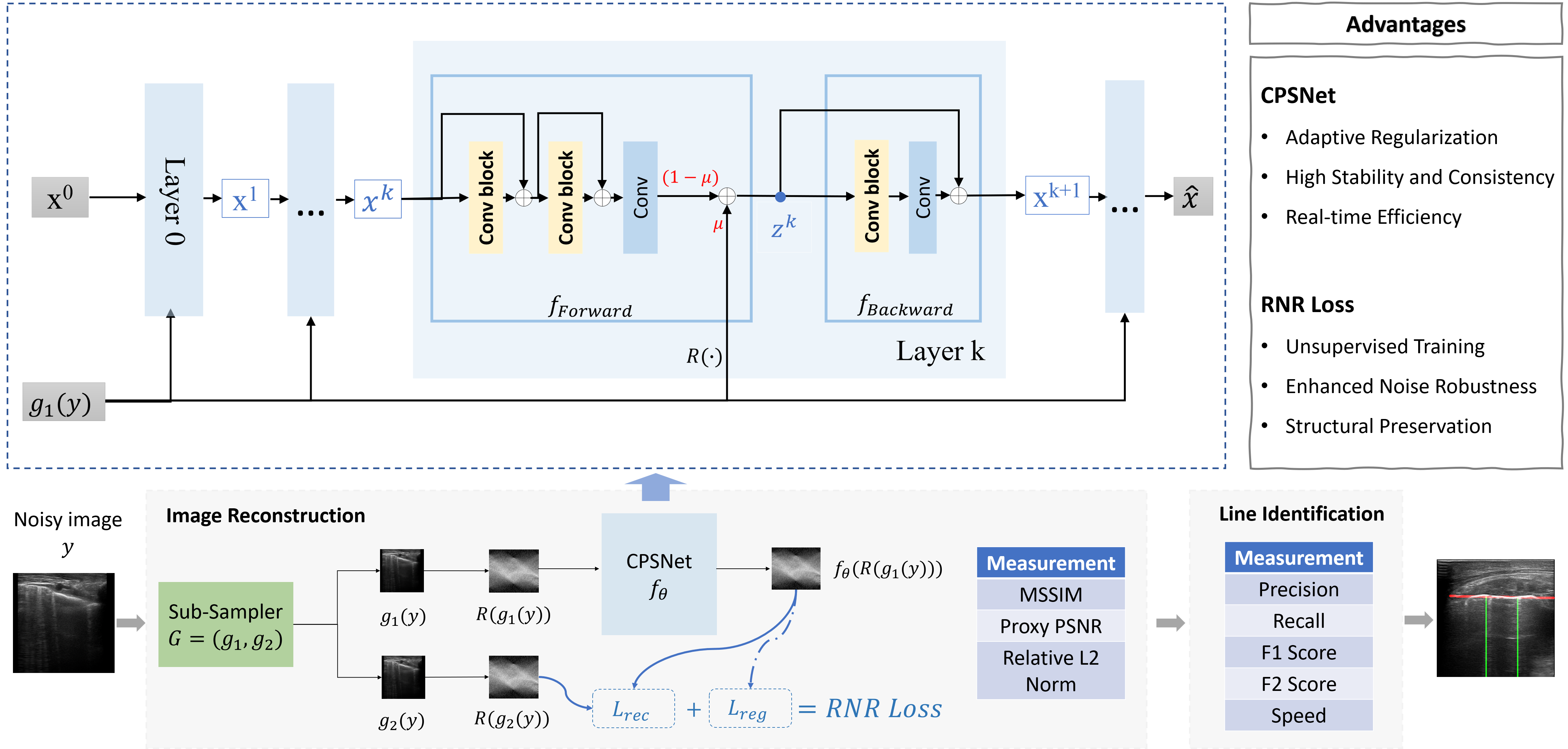}
\end{graphicalabstract}

\begin{highlights}
\item CPSNet: An unsupervised deep unfolding framework combining Cauchy proximal splitting for robust lung ultrasound image reconstruction and B-line detection.
\item Radon-Based Neighbor Reconstruction Loss (RNR Loss): Introduces a novel loss function that enables unsupervised training in the Radon domain, enhancing noise resilience and structural consistency without ground truth.
\item Adaptive Regularization with Real-Time Efficiency: CPSNet dynamically adjusts regularization to preserve critical features, achieving fast execution (0.01 seconds/frame) and stable, reliable B-line detection, making it suitable for rapid clinical applications.

\end{highlights}

\begin{abstract}
In this paper, we propose CPSNet, a label-free deep unfolding framework for lung ultrasound image analysis. 
CPSNet unfolds a Cauchy proximal splitting algorithm into a forward-backward style network architecture, incorporating skip connections to iteratively enhance noisy Radon domain images. We introduce the Radon-Based Neighbor Reconstruction Loss, a novel loss function that enforces reconstruction consistency between subsampled and reconstructed image pairs in the Radon domain, while applying regularization terms to enhance robustness against noise. Integrating the Cauchy penalty into the loss function preserves the statistical influence of the prior while enabling the network to learn flexibly.

Trained in an unsupervised manner without ground truth images, CPSNet is evaluated using structural similarity index, Proxy peak signal-to-noise Ratio, and relative L2 norm, demonstrating stable and effective performance in lung ultrasound image reconstruction. Applied to B-line detection, CPSNet achieves greater stability, adaptability and efficiency compared to traditional and object detection methods, effectively preserving line structures and minimizing false detections. This study underscores CPSNet’s potential as a reliable and efficient solution for lung ultrasound-based diagnostics, supporting more accurate and efficient clinical decision-making. The code developed for this paper is available at \url{https://github.com/TQ-001/DUCPS}.

\end{abstract}


\end{frontmatter}




\section{Introduction}
\label{sec:introduction}
Lung ultrasound (LUS) is a non-invasive and effective diagnostic tool for various pulmonary conditions, with its utility in clinical diagnostics growing significantly in recent years \cite{yang2022current}. A critical feature in LUS imaging is the presence of B-lines, which appear as vertical, hyperechoic artifacts originating from the pleural line \cite{lichtenstein2014lung,Demi2023New}. Accurate identification and quantification of B-lines can improve the diagnosis and management of lung conditions, reducing the reliance on invasive procedures like CT scans and bronchoscopy \cite{mento2022state}. 

Automated B-line detection algorithms can democratize LUS usage, allowing clinicians without specialized training to conduct reliable assessments, especially in resource-limited settings or during large-scale health crises like the COVID-19 pandemic. Moreover, understanding the formation of B-lines and their association with different pathological states offers valuable insights into pulmonary physiology and guides therapeutic interventions. For instance, B-lines can be present in early or late pneumonia \cite{Demi2023New}, and monitoring B-lines in chronic kidney disease patients can help manage fluid overload and prevent complications \cite{noble2009ultrasound, allinovi2017lung, fu2021lung}. Thus, enhancing B-line detection not only improves diagnostic accuracy but also expands the clinical applications of LUS, ensuring more patients benefit from this powerful tool.

Several techniques have been proposed to automate B-line detection in lung ultrasound. Early approaches relied on traditional image processing methods. For instance, Brattain et al. \cite{brattain2013automated} used polar reformatting and thresholding, while Moshavegh et al. \cite{moshavegh2018automatic} applied a random walk algorithm to delineate the pleural line and detect B-lines. Anantrasirichai et al. \cite{anantrasirichai2016automatic, anantrasirichai2017line} framed B-line detection as an inverse problem, transforming B-mode LUS images into a Radon-domain representation and solving it using the Alternating Direction Method of Multipliers (ADMM) \cite{boyd2011distributed}. Farahi et al. \cite{farahi2023automatic} employed wavelet denoising with Radon transforms, while Karakus et al. \cite{karakucs2020detection} utilized Cauchy proximal splitting (CPS) to regularize the solution, achieving improved performance in COVID-19 assessments. 

Deep learning methods, particularly convolutional neural networks (CNNs), have significantly improved B-line detection by automatically learning complex feature mappings. Van Sloun and Demi \cite{van2019localizing} used CNNs with gradient-based class activation mapping (Grad-CAM), and Roy et al. \cite{roy2020deep} introduced spatial transformer networks (STNs) for weakly supervised localization. The integration of domain knowledge into deep networks \cite{frank2021integrating} and domain adaptation techniques \cite{mason2021lung} further enhanced segmentation performance. Some approaches utilize both spatial and temporal information, employing 3D filters \cite{baloescu2020automated} and spatio-temporal attention mechanisms \cite{kerdegari2021b}. However, obtaining expert-annotated data remains a challenge, which has led to methods using simulated LUS images for training \cite{zhao2022covid} and contrastive learning for unsupervised representation learning \cite{yang2023semi}.

Despite these advances, model-based techniques often struggle to capture the complexity of high-dimensional data and typically demand extensive domain knowledge for accurate modeling. On the other hand, deep learning systems typically require large amounts of labeled training data, and the feature representations are difficult to intuitively understand, especially when it comes to high-level vectors. The complexity of their architectures can also lead to challenges in interpreting how individual features contribute to the final decision.  To bridge this gap, deep unfolding has emerged as an integrated approach that combines the robustness and interpretability of traditional signal processing with the adaptability of deep learning. In their seminal work, Gregor and LeCun \cite{gregor2010learning} introduced a pioneering framework called Learned ISTA (LISTA), which connects iterative algorithms like sparse coding directly with neural network architectures. This integration not only enhances performance but also improves the interpretability and understanding of underlying data for specific tasks. Building on this concept, various designs of deep unfolded networks have been explored in biomedical applications, including compressive sensing \cite{yang2018admm}, magnetic resonance imaging \cite{zhang2022high, cui2024deep}, computed tomography \cite{adler2018learned, kang2018deep, Wang2021InDuDoNetAD}, and ultrasonography \cite{van2019deep}. However, despite their demonstrated success, deep unfolded networks are predominantly trained in a supervised manner, requiring ground truth data. This dependency on labeled data limits their applicability in domains where acquiring accurate ground truth is challenging, such as B-line detection in LUS images. Addressing this limitation calls for unsupervised approaches that can leverage the advantages of deep unfolding without relying heavily on labeled datasets.

Recent studies further demonstrate that deep unfolding has become a broadly applicable paradigm for inverse imaging and vision tasks beyond conventional biomedical reconstruction. In image compressive sensing, methods such as MDGF-Net-Plus, UFC-Net, CPP-Net, NesTD-Net, USB-Net, and HUNet have incorporated wavelet-domain consistency, fixed-point continuation, primal--dual proximal point algorithms, NESTA-inspired optimization, Split Bregman iterations, and homotopy continuation into learnable network architectures, improving reconstruction accuracy while preserving algorithmic interpretability \cite{wang2024wavelet,wang2024ufc,guo2024cpp,gan2024nestd,guo2025usb,shen2025hunet}. Related inverse-imaging approaches have also explored invertible diffusion models for compressed sensing, showing that physics- or optimization-guided reconstruction remains an active research direction \cite{chen2025invertible}. Beyond compressive imaging, unfolding and model-inspired networks have been applied to low-light enhancement through Retinex-inspired reconstruction optimization \cite{zhao2024riro}, infrared small target detection through RPCA-based sparse--low-rank decomposition \cite{wu2024rpcanet}, and multi-modal MRI semantic segmentation through edge-aware robust principal component analysis \cite{zhao2025multi}. These studies highlight the growing importance of embedding explicit degradation models, proximal priors, and interpretable iterative structures into deep networks. However, most existing methods are designed for natural-image restoration, compressive sensing, or supervised medical segmentation, whereas the unsupervised detection of B-line artifacts in LUS remains less explored.

To address the above challenges, we propose CPSNet, which is specifically designed for label-free B-line detection in LUS by unfolding a Cauchy proximal splitting algorithm in the Radon domain and training the reconstruction using Radon-domain neighbour consistency, without requiring pixel-level or bounding-box B-line annotations. This network consists of a forward block to identify and enhance relevant image features, and a backward block for robust regularization, mitigating the influence of noise and preserving line structures. The Cauchy prior is applied as a global regularisation term in the loss function. By combining these mechanisms within the deep unfolding framework, CPSNet maintains the interpretability and robustness of traditional optimization while benefiting from the flexibility and feature learning capabilities of deep neural networks. A critical component of CPSNet’s unsupervised training process is the Radon-Based Neighbor Reconstruction Loss (RNR Loss) which is specifically designed to operate in the Radon domain. RNR Loss builds on the concept of Neighbor2Neighbor \cite{huang2021neighbor2neighbor} but simplifies it for Radon-domain applications. This loss function drives CPSNet to achieve consistency in the Radon domain while minimizing noise and preserving the structural integrity of B-lines.

To summarize, the main contributions of this work are:
\begin{enumerate}
    \item \textbf{CPSNet}: We propose a novel unsupervised deep unfolding framework that combines Cauchy proximal splitting for effective LUS reconstruction and B-line detection, achieving adaptive regularization and enhanced performance.
    
    \item \textbf{Radon-Based Neighbor Reconstruction Loss (RNR Loss)}: We introduce a simple yet effective loss function specifically designed for the Radon domain, enabling effective unsupervised training without reliance on ground truth data and enhancing robustness against noise.
    
    \item \textbf{Adaptive Regularization and Improved Stability}: CPSNet’s deep unfolding architecture dynamically balances noise suppression with structural preservation, avoiding over-regularization and maintaining essential image features, achieving more consistent results than traditional and object detection methods.
    
    \item \textbf{High Efficiency for Real-Time Applications}: CPSNet demonstrates rapid execution, making it suitable for real-time clinical applications. This efficiency, combined with stable performance, supports timely diagnostic assessments and complements high-demand clinical workflows.

\end{enumerate}

The remainder of this paper is organized as follows: Section \ref{sec:DUCPS} details the architecture of CPSNet and the construction of the proposed RNR Loss. Section \ref{sec:exp} describes the datasets, evaluation metrics, and training details used to evaluate CPSNet’s performance and the application to B-line detection. In Section \ref{sec:results}, we present the experimental results, comparing CPSNet with traditional algorithms and popular object detection neural networks. Finally, we summarize the findings of the paper in Section \ref{sec:conclusion} and discuss potential directions for future work.

\section{Methodology}
\label{sec:DUCPS}

\begin{figure*}[t!]
\centering
\includegraphics[width=\linewidth]{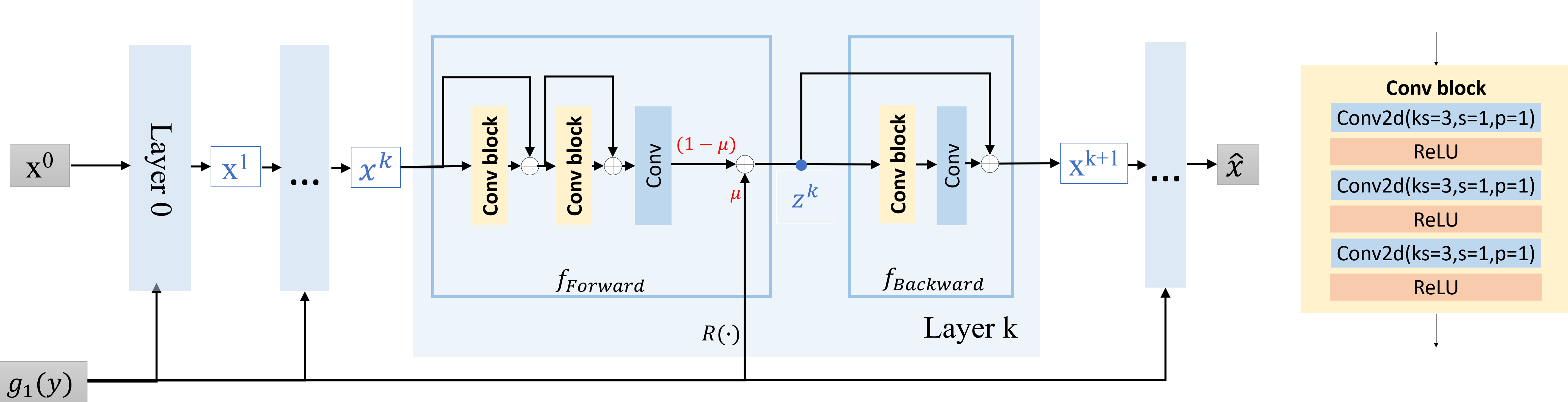}
\caption{Overall architecture of the proposed CPSNet. The network is designed as a deep unfolding of the forward–backward optimisation algorithm, with similar structures repeated across multiple layers. For illustration, the diagram shows a representative layer $k$, consisting of two main components: the Forward Block, which performs data-consistency updates, and the Backward Block, which refines the intermediate estimate using learned feature transformations. Input, intermediate, and output variables are shown in grey and trainable parameters in red. The detailed design of the Conv block is provided on the right. The convolutional layers use $kernel\_size=3$, $stride=1$, and $padding=1$.}
\label{fig:CPSNet}
\end{figure*}

\subsection{Line artefact model}
\label{sec:model}
In LUS analysis, one of the primary objectives is to accurately represent and detect line structures such as B-lines, which appear as resonance artifacts in ultrasound images \cite{Demi2023New}. To achieve this, we write the inverse problem in the Radon domain as follows:
\begin{equation}\label{model}
y=R^{-1}x+n,
\end{equation}
where $y$ denotes the observed noisy spatial image, $x$ is the set of lines or features in the Radon domain, $R^{-1}$ is the inverse Radon transform operator, which maps the Radon domain back to the spatial domain, $n$ signifies additive white Gaussian noise (AWGN). We employ the Radon transform due to its robust capabilities in detecting and enhancing linear features, especially within noisy environments typical of LUS images. This technique excels at converting spatial lines into distinct peaks in the parameter space, significantly aiding in the identification of linear structures like B-lines. Moreover, the Radon transform's integration over straight lines naturally suppresses random noise, enhancing both the accuracy and reliability of line detection in our proposed model.

The goal is to estimate the set of line structures $x$ from the observed image $y$. This estimation is challenging due to the presence of noise and the complexity of mapping between the Radon domain and the spatial domain. Traditionally, this inverse problem can be solved using iterative optimization algorithms refining the estimate of $x$ by minimizing a combination of a data fidelity term and a regularization term. 

In particular, the Cauchy regularization term \cite{karakucs2020detection} has been effectively used to handle outliers and enforce sparsity in the reconstruction. The use of the Cauchy prior is motivated by the expected structure of B-line representations in the Radon domain, where true line artefacts correspond to sparse, high-amplitude peaks embedded in weaker background responses. Compared with quadratic regularization, which tends to over-smooth salient structures, the Cauchy penalty provides a heavy-tailed prior that is more robust to outliers and preserves strong Radon-domain responses. Compared with standard ($\ell_1$)-type sparsity regularization, the Cauchy prior imposes weaker shrinkage on large coefficients, thereby reducing amplitude bias and better retaining prominent line features. These properties make it suitable for modelling sparse line artefacts in noisy LUS images while suppressing background fluctuations.

The traditional methods often require manual tuning of parameters and are limited in their ability to learn complex mappings from noisy observations to clean representations. To address these limitations, we propose CPSNet, a novel deep learning framework that unfolds the traditional forward-backward optimization algorithm into a neural network. This approach leverages the interpretability and robustness of traditional model-based methods while incorporating the flexibility and learning capabilities of deep neural networks.

\subsection{CPSNet}
\label{sec:CPSNet}
The architecture of CPSNet is designed to iteratively refine the estimated line representation in the Radon domain. As shown in Figure \ref{fig:CPSNet}, CPSNet consists of two primary components: the Forward Block and the Backward Block, which operate in a layered structure. 

\subsubsection{Forward Block}The Forward Block, denoted as $f_{Forward}$, is designed to generate an intermediate representation $z^{k}$ using the current estimate $x^{k}$ and the noisy observation $y$. The mathematical expression can be shown as

\begin{align}\label{Forward}
z^{k} & = x^{k}-\mu(R^{-1})^{T}(R^{-1}x^{k}-y) \\
& = f_{Forward}(x^{k},y)
\end{align}.

The network begins with a series of conv blocks, each consisting of alternating convolutional layers and ReLU activations. These blocks are designed to extract multi-scale features from the input estimate $x^{k}$, enabling the network to capture complex patterns. Skip connections are incorporated to enable the input $x^{k}$ to directly influence the output through a weighted summation, expressed as: 
\begin{equation}
    z^{k}=(1-\mu)\cdot \mathrm{Convs}(x^{k},y)+\mu\cdot R(y)
\end{equation}
Here, $\mu$ is a learnable parameter that dynamically balances the contributions from the learned structures and the original input, ensuring an adaptive trade-off between preserving input features and leveraging new information to refine the output. 

\subsubsection{Backward Block} 
The Backward Block, denoted as $f_{Backward}$, is responsible for refining the intermediate representation $z^{k}$  obtained from the forward step, producing the updated estimate $x^{k+1}$. In the classical forward–backward algorithm, this stage corresponds to applying the proximal operator of the regularisation term. In our formulation, it is chosen as the negative log-likelihood of a Cauchy prior which promotes sparsity and robustness through its heavy-tailed distribution. 
\begin{equation}
    P_{Cauchy}(x) = -\log{\frac{\gamma }{\gamma ^{2}+u^{2}} },
\end{equation}
where $\gamma$ is the scale parameter, controlling the spread of the Cauchy distribution. 

Theoretically, the backward step would involve computing the exact Cauchy proximal mapping. However, two practical challenges arise. One is the intractability in high-dimensional convolutional spaces – while the scalar form of the Cauchy proximal operator can be derived analytically, extending it to multi-channel feature maps with spatial correlations is not straightforward. The other is the loss of learning flexibility – embedding a fixed, closed-form proximal operator inside the network layers would constrain the representational capacity of the backward step, potentially limiting its ability to adapt to the complex, data-dependent structures of LUS Radon-domain signals.
     
To address these issues, we adopt a hybrid strategy:
\begin{enumerate}
 \item The Backward Block is implemented as a fully learnable feature refinement module composed of stacked Conv–ReLU blocks and a residual connection. The convolutional layers capture local context and enhance feature representations, while the residual connection stabilises training and prevents excessive alterations to the intermediate estimate.
 \item The Cauchy prior is applied as a global regularisation term in the loss function, rather than being explicitly embedded in the layer operations. This preserves the statistical influence of the prior across the final network output while allowing the backward step to remain a flexible, data-driven approximation of the proximal mapping.
\end{enumerate}

This design choice maintains consistency with the forward–backward paradigm — where the backward step still plays the role of regularisation — while leveraging the strengths of deep learning for feature adaptation. In effect, the learnable Backward Block approximates the behaviour of a Cauchy proximal operator, and the loss-level regularisation guides the output toward solutions that embody the intended heavy-tailed prior.

\subsection{Radon-Based Neighbor Reconstruction Loss}
\label{sec:RNRLoss}

\begin{figure*}[t!]
\centering
\includegraphics[width=\linewidth]{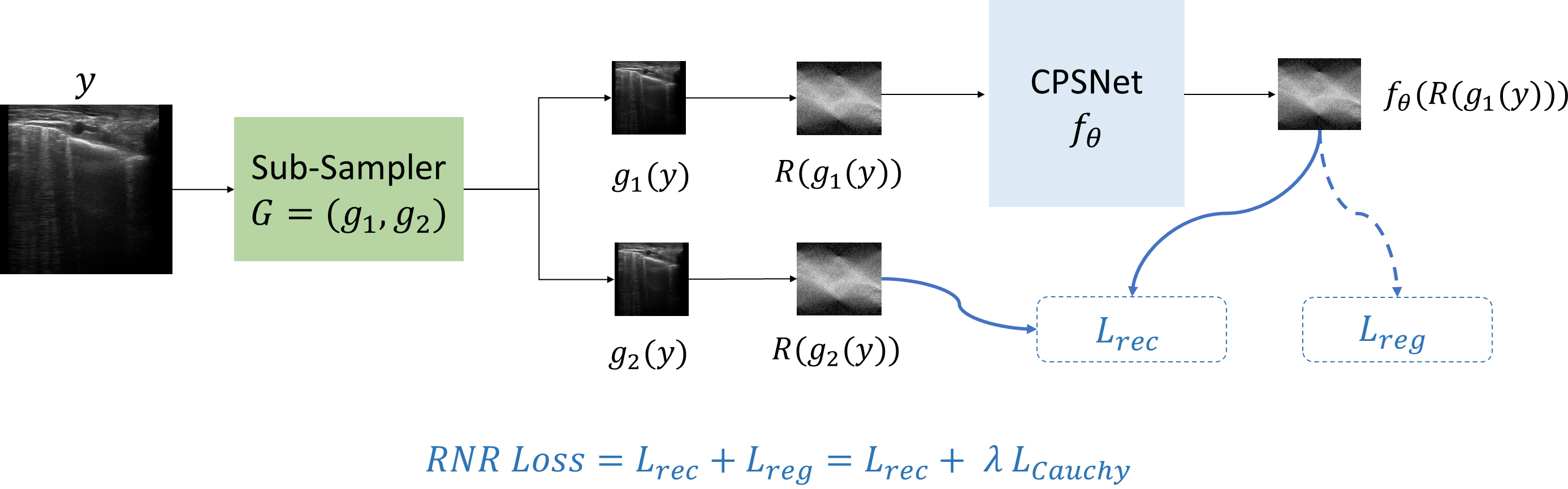}
\caption{The overall pipeline of CPSNet, training with RNR Loss.}
\label{fig:RNRLoss}
\end{figure*}

The success of CPSNet depends on a loss function that aligns with its unsupervised training strategy and operates in the Radon domain, effectively saving the computation load of performing Radon transformation in each iteration. To this end, we propose the Radon-Based Neighbor Reconstruction Loss (RNR Loss), a simple yet powerful loss function that leverages the relationships between paired noisy images in the Radon domain. Inspired by the Neighbor2Neighbor approach \cite{huang2021neighbor2neighbor}, RNR Loss is designed to enforce consistency between subsampled radon image pairs and incorporate regularization terms that enhance noise suppression and structural preservation. 

\subsubsection{Motivation for RNR Loss}
The RNR Loss draws theoretical justification from the principles of the Noise2Noise method \cite{lehtinen2018noise2noise}, where deep neural networks are trained using only noisy images. In Noise2Noise, the noise is assumed to have a zero mean, but independence or uncorrelated properties of two realizations were not strictly required. Empirically, however, independent noise realizations were shown to be effective \cite{zhussip2019extending}. In CPSNet, we follow a similar approach, generating two similar images with independent noise using a neighbor subsampler \cite{huang2021neighbor2neighbor}. These subsampled images serve as the input for Radon transformation and network training.

Since the noise in the spatial domain $n$ is assumed to be zero-mean, the linearity of the Radon transform ensures that the zero-mean property is preserved in the Radon domain. This is a common property for AWGN. Mathematically, if $E[n]=0$, then $E[R(n)]=0$ holds, where $R(\cdot)$ denotes the Radon transform. This zero-mean assumption underpins the reconstruction loss calculation and allows CPSNet to learn meaningful features in the Radon domain without requiring ground truth images.

\subsubsection{Loss Definition}
As shown in Figure \ref{fig:RNRLoss}, the RNR Loss is composed of three main steps: neighbor subsampling, Radon transformation, and loss computation.

\textbf{Neighbor Subsampling}: Given a LUS image $y$, we apply a neighbor subsampler to generate two subsampled spatial images $g_{1}(y)$ and $g_{2}(y)$. This subsampling strategy involves selecting neighboring pixels from each $2 \times 2$ patch in $y$ to create $g_{1}(y)$ and $g_{2}(y)$, resulting in two images that retain spatial correlations while being independently corrupted by noise.

\textbf{Radon Transform}: We apply the Radon transform to the subsampled images, yielding $R(g_{1}(y))$ and $R(g_{2}(y))$ in the Radon domain. One of the transformed images will be refined by the deep unfolded network to facilitate line representation, which is critical for accurate B-line detection.

\textbf{Loss Computation}:  The RNR Loss is a weighted combination of the reconstruction loss and the regularization term.
\begin{enumerate}
\item Reconstruction Loss $L_{rec}$: CPSNet reconstructs $R(g_{1}(y))$ as $f_{\theta}(R(g_{1}(y)))$, where $\theta$ represents the set of parameters of the network. The reconstruction loss is then defined as the mean squared error (MSE) between the reconstructed image $f_{\theta}(R(g_{1}(y)))$ and the paired noisy image $R(g_{2}(y))$:
\begin{equation}
    L_{rec}=\mathrm{MSE} (f_{\theta}(R(g_{1}(y))),R(g_{2}(y)))
\end{equation}
This loss term ensures consistency between the reconstruction and the paired input in the Radon domain, effectively reducing random noise while preserving important structures, rather than relying solely on a single noisy observation.
\item Regularization Term $L_{reg}$: The Cauchy penalty suppresses outliers by leveraging the properties of the Cauchy distribution, enforcing sparsity in the reconstruction:
\begin{equation}
    L_{Cauchy}=\sum _{i,j}\log{(1+\gamma \cdot (f_{\theta}(R(g_{1}(y))) _{i,j})^2)} 
\end{equation}
where $\gamma$ is the parameter that controls the weight or influence of large values in the regularization term. The indices $i,j$ represent the spatial coordinates of the image or feature map, indicating that the regularization term is applied element-wise across all pixels in the output $f_{\theta}(R(g_{1}(y)))$.
\end{enumerate}
Therefore, the final RNR Loss is 
\begin{equation}
    L_{RNR}=L_{rec} + L_{reg} = L_{rec}+\lambda \cdot L_{Cauchy}
\end{equation}
where $\lambda$ is the hyperparameter that controls the importance of the penalty component, and is determined empirically to achieve a balance between noise suppression and structural preservation.

\subsection{B-line Identification}

\begin{figure*}[t!]
\centering
\begin{subfigure}[t]{\textwidth}
        \centering
        \includegraphics[width=\textwidth]{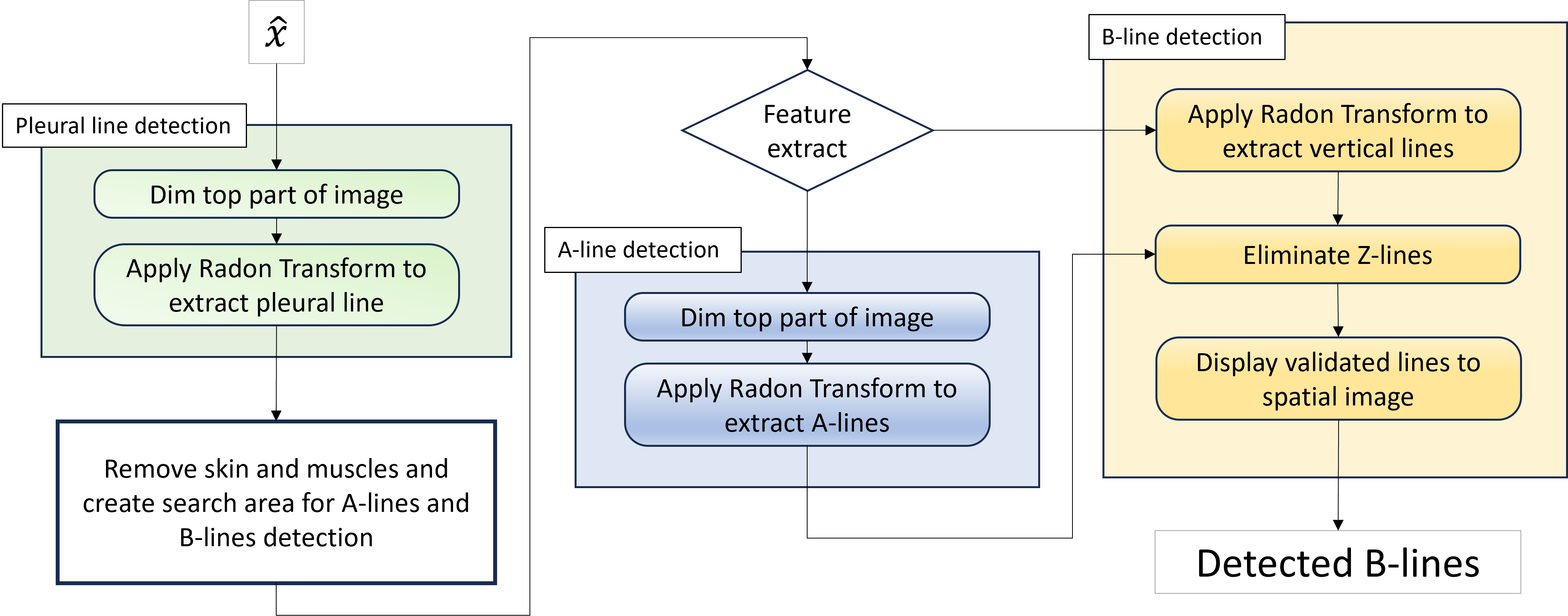}
        \caption{The framework to extract lung features including pleural line, A-lines and B-lines \cite{anantrasirichai2017line}.}
        \label{fig:linedetectionsteps}
    \end{subfigure}
    \hfill
    \begin{subfigure}[t]{\textwidth}
        \centering
        \includegraphics[width=\textwidth]{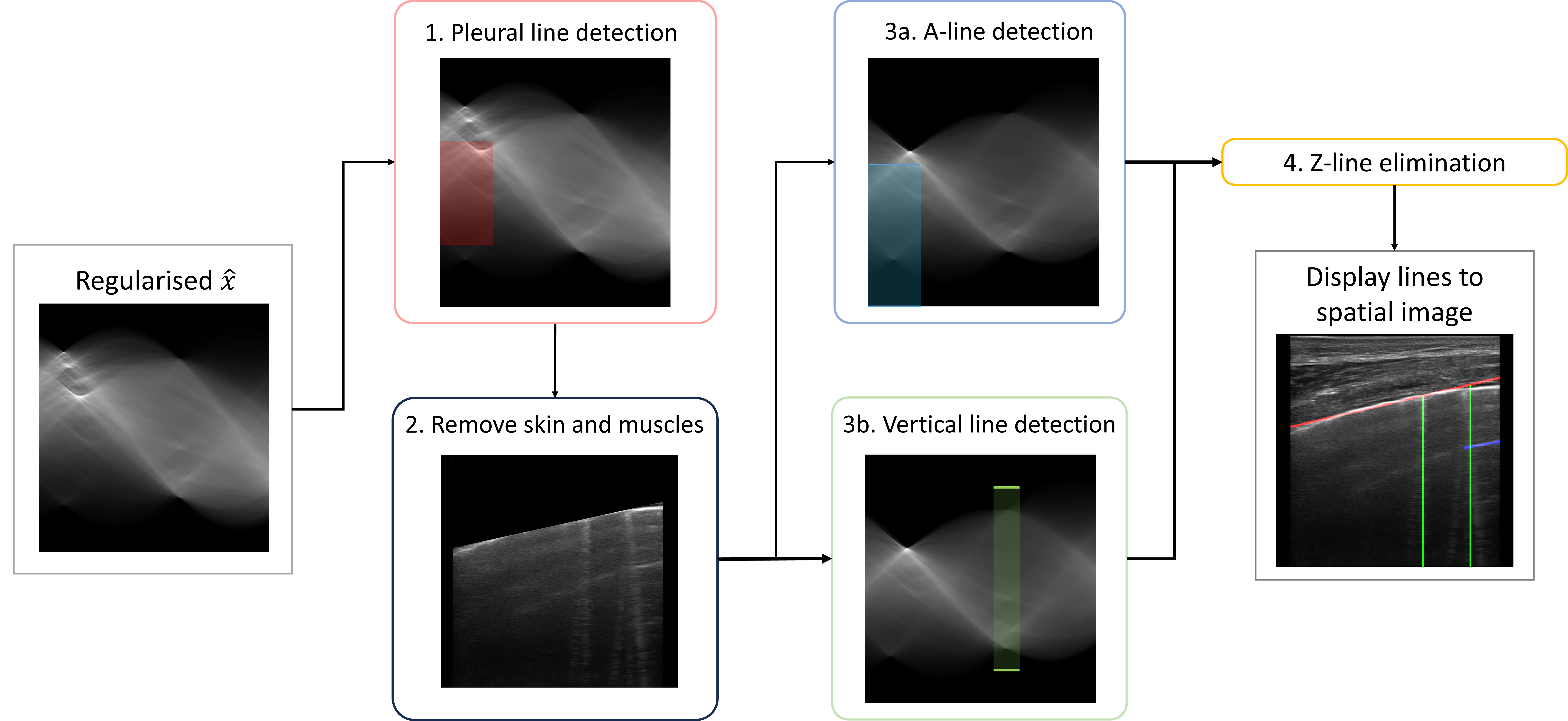}
        \caption{The process of identifying line artifacts. Searching areas for the pleural line (red), A-line (blue), and B-line (green) are masked in the Radon domain. The detected pleural line is displayed in red, the A-line is in blue, and the B-lines are depicted in green.}
        \label{fig:lineidentification}
    \end{subfigure}
    \caption{Label-free B-line identification procedure in the Radon domain.}
\end{figure*}

The line identification steps are illustrated in Figure \ref{fig:linedetectionsteps} and \ref{fig:lineidentification}. The procedure is implemented in the Radon domain using the local maxima technique. It is crucial to set different searching areas to distinguish line artifacts considering that they possess characteristic structures in the spatial domain. This includes limiting the searching angles and upper and lower borders in the Radon domain. For pleural line and A-lines, we limit the searching angle within the range $\theta_h \in [90^{\circ} \pm 5^{\circ} ]$ ($0^{\circ}$ originates from the $x$ axis, which indicates a horizontal line) as they are generally horizontal. Similarly, the B-lines are detected at $\theta_v \in [ \pm15^{\circ}]$ as they are mainly vertical.

\section{Experimental Setup}
\label{sec:exp}
In this section, we describe the datasets used for training and evaluating CPSNet, and the evaluation metrics employed to measure its performance regarding image reconstruction. In the application of B-line identification, we followed the procedure in \cite{anantrasirichai2017line}, then introduced a scoring rule to fairly evaluate the performance of the methods. Finally, we describe the details of the method implementation for running the experiments.

\subsection{Dataset}
\label{sec:data}
The ultrasound data, in the form of image sequences, were obtained from 45 patients from the Nephrology, Dialysis and Transplantation Department, Careggi University Hospital, Florence, Italy. LUS evaluations were performed whilst patients attended for regular hemodialysis, using an ultrasound machine (MyLab Class C-Esaote$^{\circledR}$, Genoa, Italy) with a 6–18 MHz linear probe. The data was saved in DICOM format.

Data preprocessing includes extracting the frames from the video clips, cropping the region that only contains lung information, and finally zero-padding all the images to the size of 480$\times$600 pixels (height $\times$ width).

\begin{figure*}[t!]
\centering
\includegraphics[width=\linewidth]{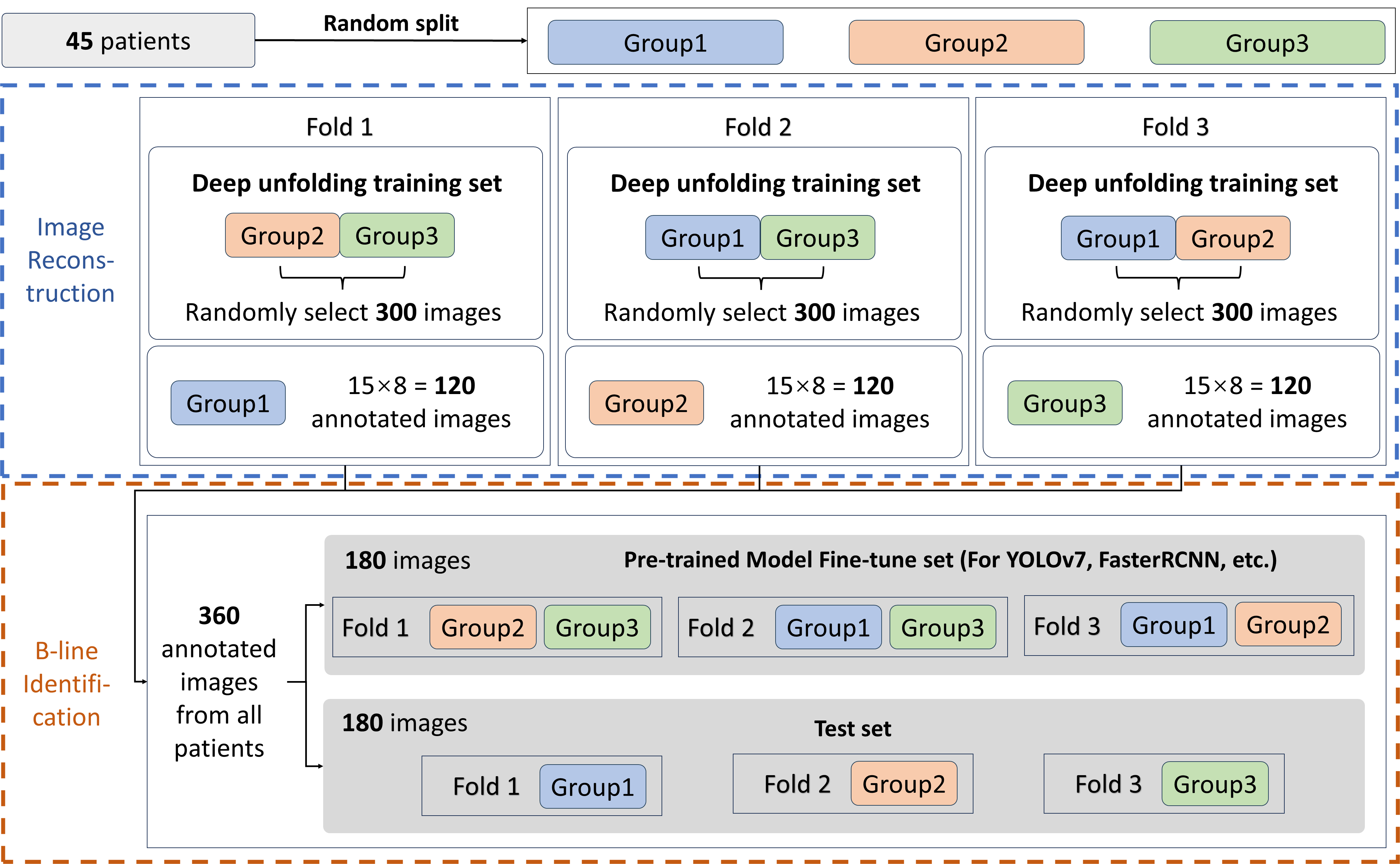}
\caption{Illustration of setting up the dataset. Data random splitting is implemented twice, each of which is designed to perform a 3-fold validation.}
\label{fig:dataset}
\end{figure*}

The detailed data splitting strategy is illustrated in Figure~\ref{fig:dataset}. We randomly split the dataset into three groups (Group1, Group2, and Group3) at the patient level. Each of the groups contains data from 15 patients. By partitioning the dataset into 3 subsets, we evaluate the proposed method in a 3-fold manner. This helps prevent information leakage between groups and iteratively uses each subset for testing while the remaining subsets are used for training. The training set in each fold is formed by randomly selecting 300 images from two of the groups (for example Group1 and Group2). Then we randomly select 8 images from each patient in the remaining group (for example Group3), and a physician with long-term expertise in LUS provides the annotation of the line artefacts as the ground truth. The annotated images are subsequently divided into two sets which are used for pre-trained model (for example YOLO \cite{wang2023yolov7, yolo11_ultralytics} and FasterRCNN \cite{ren2015faster}) fine-tuning and algorithm testing respectively. The final test set contains in total 180 images. The whole procedure, starting from random splitting, is implemented twice to enhance result stability and increase the reliability and credibility of the experiment by assessing consistency and minimizing uncertainty.

\subsection{Evaluation Metrics}
\label{sec:metrics}
\textbf{Image Reconstruction}: Common metrics to quantify a reconstructed image include Structural Similarity Index (SSIM) \cite{wang2004image}, Peak Signal-to-Noise Ratio (PSNR), and L2 Norm. The SSIM measures the structural similarity between the reconstructed image and the reference noisy image. It captures the perceptual quality of the reconstructed image, focusing on luminance, contrast, and structural information. A higher SSIM value indicates better structural preservation. In this work, we use the mean SSIM (MSSIM) over a batch.  
The key challenge for evaluating the reconstruction performance of CPSNet is that we do not have access to clean
images, and therefore the standard PSNR cannot be calculated directly in our unsupervised setting. To provide a comparable measure, we introduce a Proxy PSNR defined as 
\begin{equation}
\begin{split}
    \text{Proxy PSNR} &= 10 \log_{10} \left( \frac{\text{MAX}_I^2}{\frac{1}{N} \| x_{\text{input}} - x_{\text{recon}} \|_2^2} \right),
\end{split}
\end{equation}
where $\text{MAX}_I$ is the maximum pixel value of the image (in our case is 1) and $N$ is the total number of pixels. It measures relative deviation from the surrogate reference, not absolute fidelity to the clean image. A higher value indicates that more of the original signal has been preserved, with lower residual noise. 

Similarly, we define a relative L2 norm to measure how closely the network’s output resembles the noisy input:
\begin{equation}
\text{Relative L2 Norm} = \frac{\| x_{\text{output}} - x_{\text{input}} \|_2}{\| x_{\text{input}} \|_2}.
\end{equation}
Although the network’s goal is to denoise or regularize the image, we want to avoid overly aggressive transformations that diverge too much from the input. A lower relative L2 norm indicates that the network’s output remains close to the input while reducing noise.

\begin{figure}[t!]
\centering
\includegraphics[width=0.5\linewidth]{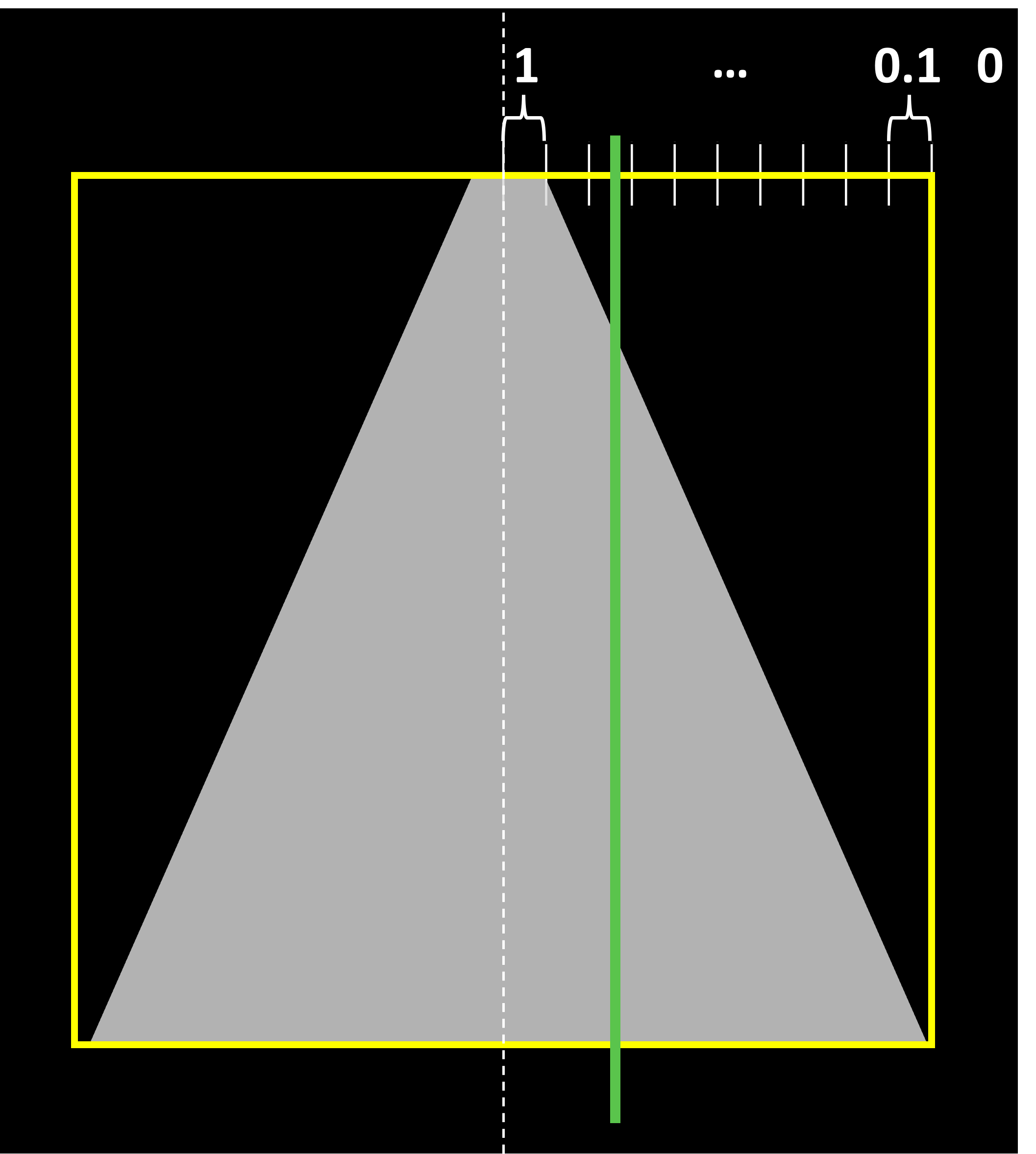}
\caption{An example of scoring the detected B-line (gray area). In this case, the green line is scored 0.8.}
\label{fig:score}
\end{figure}
\textbf{B-line Identification}: To statistically evaluate the performance in B-line detection, we proposed a scoring rule. The scoring system is designed to be scale-independent to allow for flexibility across different scales of B-line visualizations. It provides a symmetrical and graded assessment of detection accuracy, offering precise feedback on the positioning of detections relative to the target area. As sketched in Figure \ref{fig:score}, the yellow bounding box (bbox) represents the ground truth where a B-line is located, and the green line indicates the location of detection. The width of the bbox is divided into ten regions on each side of a center line. The closer the green line is to the center line, the higher the score it receives, with scores ranging from 1 to 0.1 in decrements of 0.1 as the distance from the center line increases. Each region closest to the center line scores 1, decreasing by 0.1 for each subsequent region moving away from the center. Any green lines falling outside the bbox receive a score of 0, indicating no accuracy in detection within the target area. 

Notably, mapping the detected B-lines to their corresponding bounding boxes presents several challenges. (i) \textbf{1 vs N}: This scenario occurs when a single detected B-line falls in the overlap of multiple ground truth bboxes. (ii) \textbf{N vs 1}: This case arises when multiple detected B-lines fall within a single ground truth bbox. To address these issues, we implemented a two-way validation strategy. In the first case (1 vs N), we computed scores for each detected line with respect to each overlapping bbox and assigned the detection to the bbox with the highest score, ensuring that each detected line is counted only once. In the second case (N vs 1),  we considered all the detections within one bbox and retained only the detection with the highest score, ensuring that each ground truth bbox is matched to a single detected B-line. By validating the detections in both directions, we ensure an accurate and reliable mapping between the predicted and ground truth annotations.

After identifying the B-lines, metrics including precision, recall, F1 and F2 score are used to provide a comprehensive assessment of CPSNet’s effectiveness in detecting B-lines, complementing the evaluation metrics used for image reconstruction. Threshold-based proxy mAP was computed by sweeping the score threshold from 0 to 1, treating detections with scores above each threshold as true positives, deriving the corresponding false positives and false negatives from the total detections and ground-truth counts, and integrating the resulting precision--recall curve.

\subsection{Implementation Details}
\label{sec:implement}

\subsubsection{Training Configuration and Complexity}
The proposed CPSNet is implemented using PyTorch and runs on an NVIDIA GeForce RTX 3090 GPU. The Radon transform is performed over a 180-degree range, spanning from $-110^{\circ}$ to $70^{\circ}$ ($0^{\circ}$ starts from the $x$ axis) with a step interval of $0.5^{\circ}$. We use the Adam optimizer with the initial learning rate of $10^{-4}$. A learning rate scheduler is employed to reduce the learning rate by a factor of 0.5 after every 5 epochs. The model is trained for a total of 20 epochs with a training batch size of 4 and a validation batch size of 1. The learning parameter $\mu$ is initialized to be 0.5. In the RNR Loss, the regularization parameter $\lambda$ is set to 0.01, while the parameter $\gamma$ in the Cauchy penalty term is set to 0.5. During the ablation study, we explore the impact of the number of layers in CPSNet, ranging from 1 to 10 layers, and the influence of the regularization in the loss function.

The computational complexity of CPSNet is mainly determined by the convolutional operations in each unfolded layer. Let the Radon-domain input size be $H \times W$, and let $K$ denote the number of unfolded CPS layers. In each layer, the forward sub-network requires approximately
\begin{equation}
9HW(1\times32 + 5\times32\times32 + 32\times1)=46656HW,
\end{equation}
multiply--accumulate operations (MACs), while the Cauchy proximal sub-network requires approximately
\begin{equation}
9HW(1\times64 + 2\times64\times64 + 64\times1)=74880HW,
\end{equation}
MACs. Therefore, the total computational cost is approximately $121536KHW$ MACs, excluding the negligible cost of scalar fusion and activation functions. For the 10-layer CPSNet used in this study, the complexity is approximately $1215360HW$ MACs. Since the same CPS block is reused across unfolded iterations, the number of trainable parameters does not increase with $K$, resulting in approximately $1.22 \times 10^5$ trainable parameters.

\subsubsection{Evaluation}
To evaluate the performance of CPSNet in B-line detection, we compare it with several baseline methods, including PUI \cite{anantrasirichai2017line}, CPS \cite{karakucs2020detection}, DUBLINE \cite{yang2023dubline}, YOLOv7 \cite{wang2023yolov7}, YOLOv11 \cite{yolo11_ultralytics} and FasterRCNN \cite{ren2015faster}. For model-based methods, we used MATLAB R2023a with 64-bit OS 12th Gen Intel$^{\circledR}$ Core\textsuperscript{TM} i7-12700 2.10 GHz. The algorithm was executed on a CPU, as the problem size and operations did not offer meaningful acceleration on a GPU. DUBLINE is tested on the same device as CPSNet, while YOLOv7, YOLOv11 and FasterRCNN are fine-tuned with early stopping and executed on the same GPU for a fair comparison. 

In the proposed scoring rule, a threshold of 0.5 was strategically selected to distinguish true detections from false ones, providing a balanced criterion that considers both precision and practical detectability in a clinical setting. Detections with scores below 0.5 are classified as false positives, while those exceeding the threshold are considered true positives. This threshold ensures an optimal balance, preventing an overly lenient approach that may allow inaccurate localizations while also avoiding excessive strictness that could exclude clinically relevant detections that are slightly off-center but still diagnostically meaningful.

For YOLOv7, YOLOv11 and FasterRCNN, the confidence thresholds varied slightly across folds to maximize F1 score, which is expected due to differences in training subsets. This ensures an unbiased estimate of generalization performance and reflects the fact that the optimal operating point may vary with the training data distribution in each fold.

\section{Results Analysis} 
\label{sec:results}
In this section, we evaluate the performance of CPSNet in both image reconstruction and B-line identification. Additionally, we conduct ablation studies to assess the effectiveness of the proposed schemes. The evaluation is performed using both subjective and objective analyses, based on the metrics introduced earlier.

\subsection{Image Reconstruction}
\label{sec:reconstruction}
\begin{figure}[t!]
    \centering
    \begin{subfigure}[b]{0.49\textwidth}
        \centering
        \includegraphics[width=\textwidth]{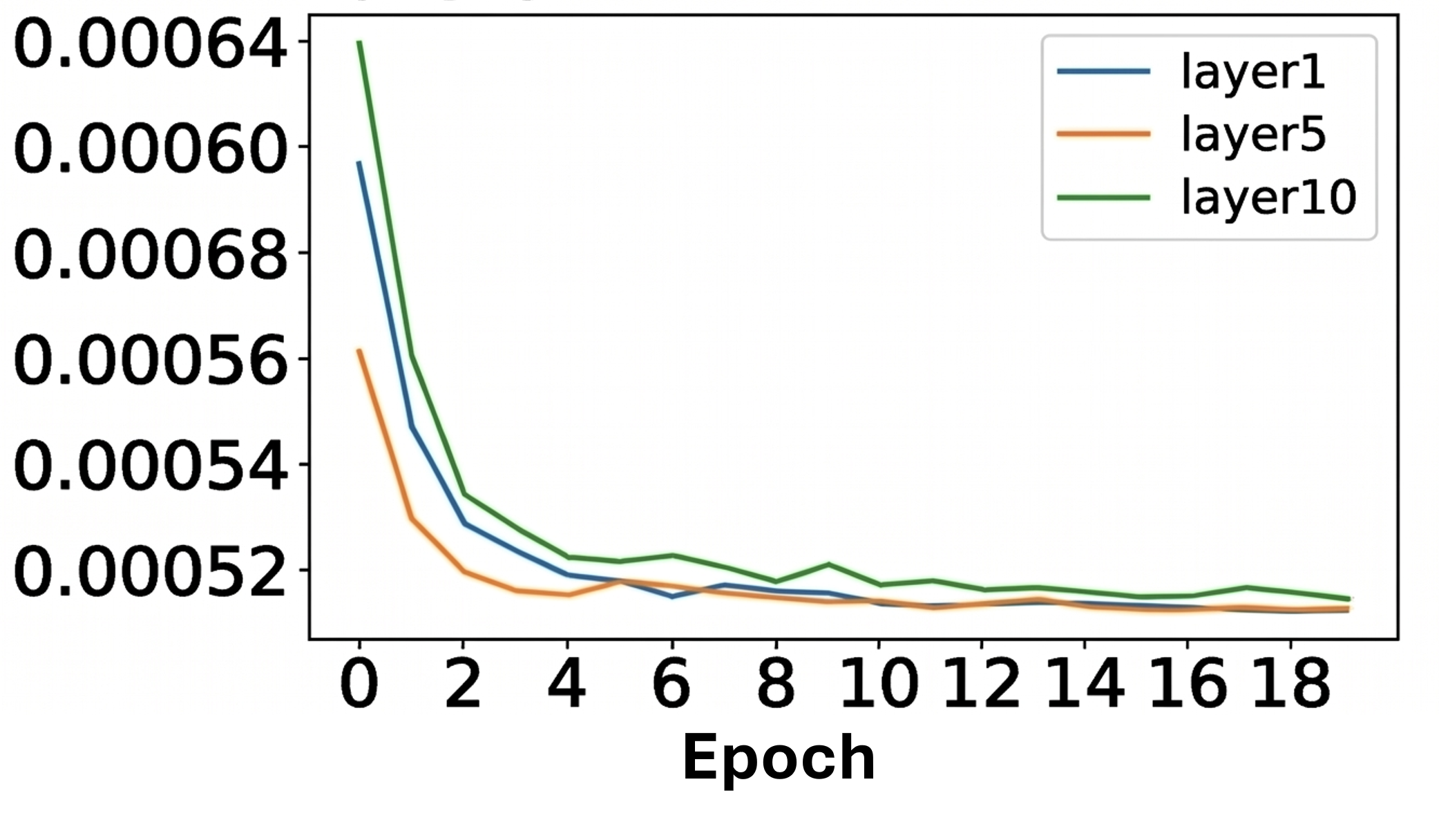}
        \caption{Validation Loss}
        \label{fig:val_loss}
    \end{subfigure}
    \hfill
    \begin{subfigure}[b]{0.49\textwidth}
        \centering
        \includegraphics[width=\textwidth]{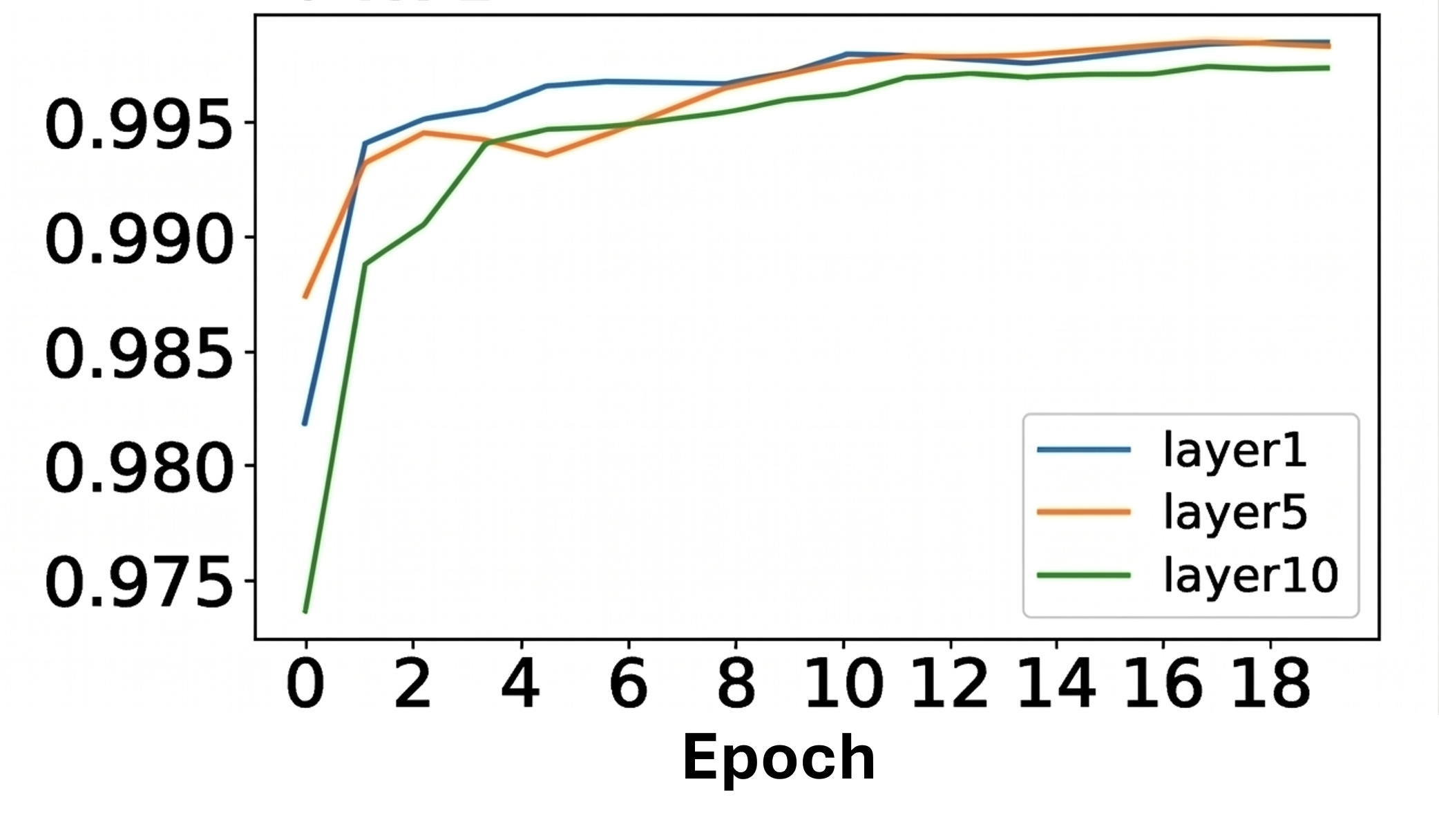}
        \caption{Validation MSSIM}
        \label{fig:val_mssim}
    \end{subfigure}

    \begin{subfigure}[b]{0.49\textwidth}
        \centering
        \includegraphics[width=\textwidth]{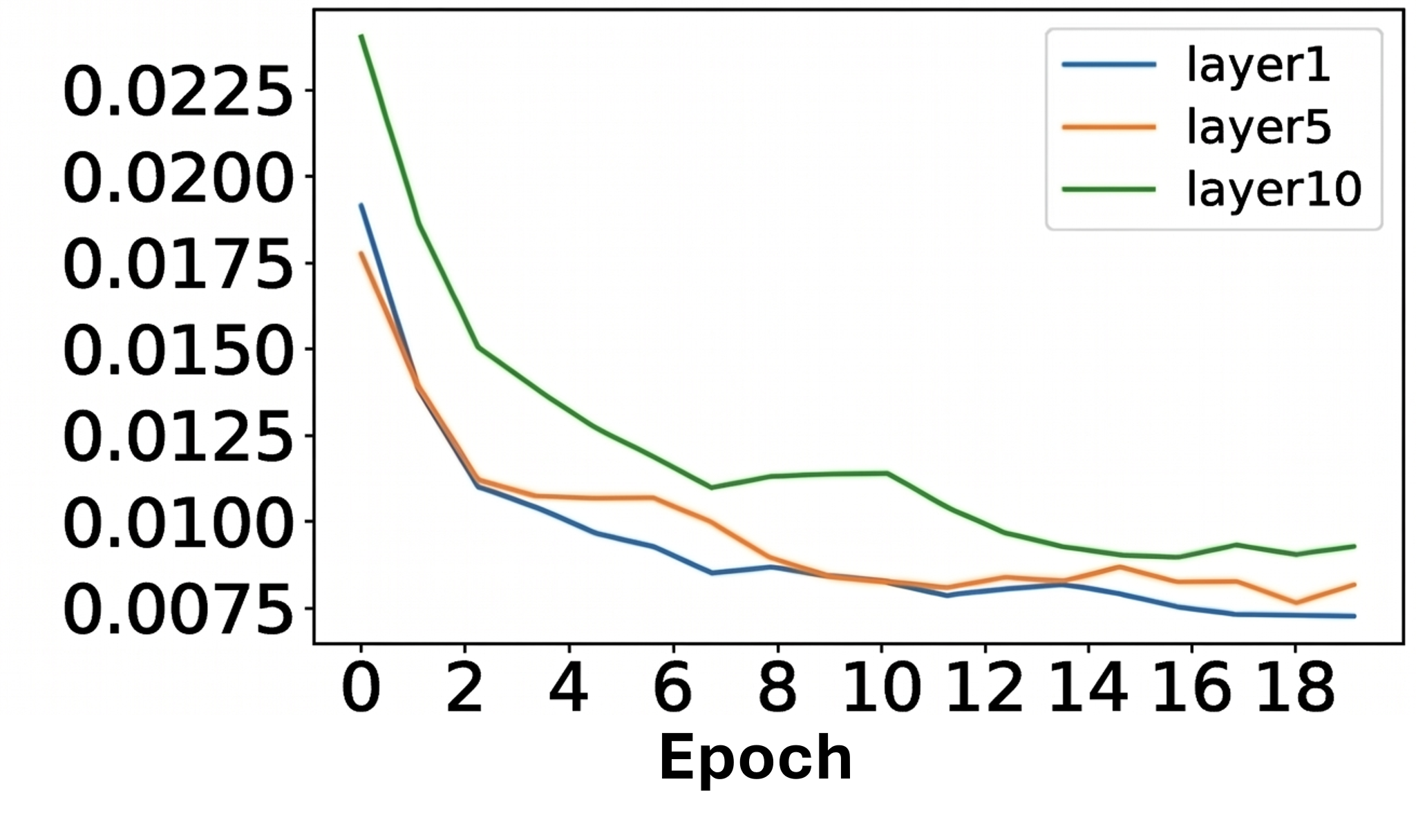}
        \caption{Validation Relative L2 Norm}
        \label{fig:val_l2}
    \end{subfigure}
    \hfill
    \begin{subfigure}[b]{0.49\textwidth}
        \centering
        \includegraphics[width=\textwidth]{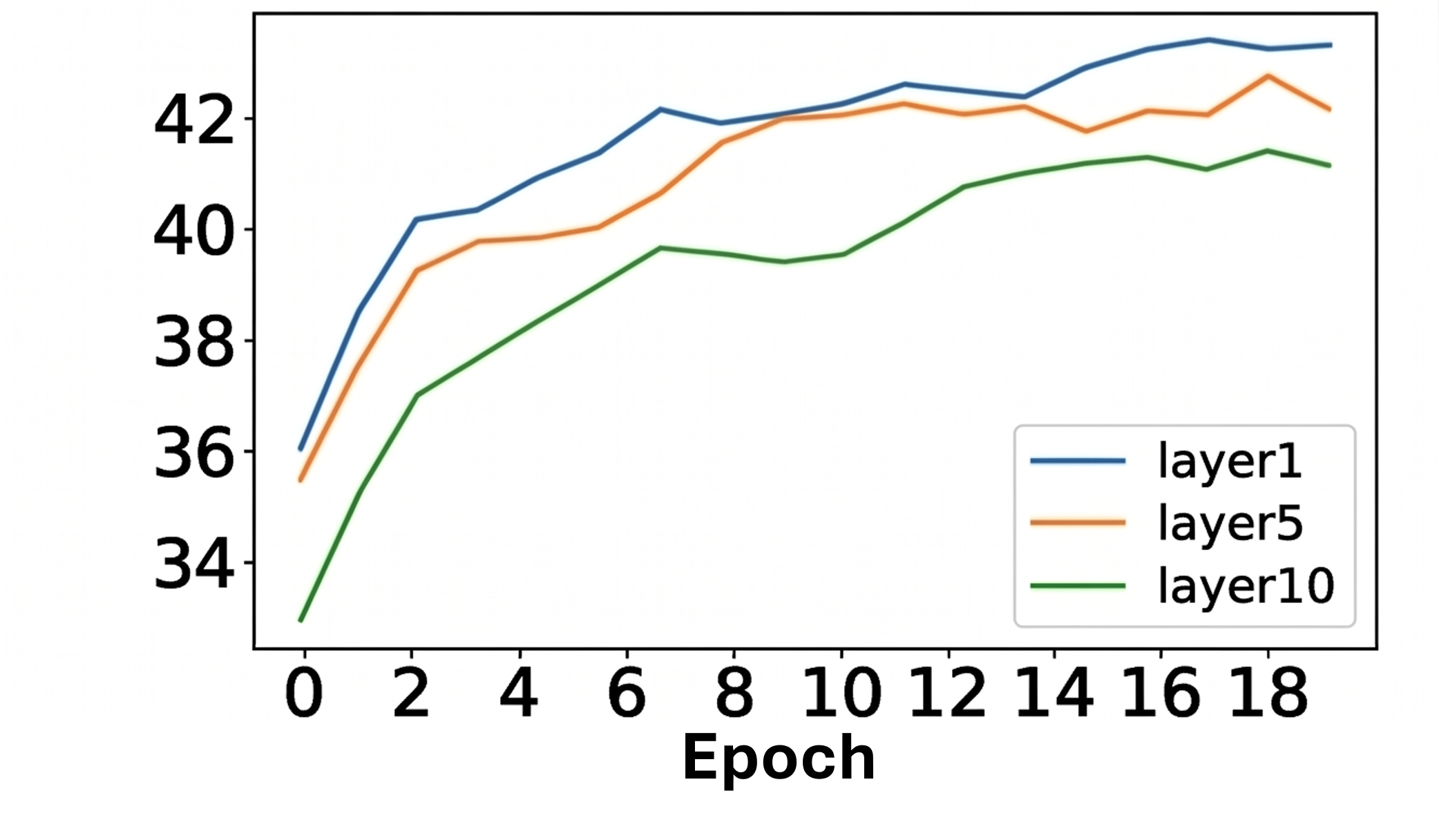}
        \caption{Validation Proxy PSNR}
        \label{fig:val_SNR}
    \end{subfigure}

    \caption{Performance evaluation of CPSNet using validation metrics. For clarity of visualization, only representative curves corresponding to networks with 1, 5, and 10 layers are shown in the figure.}
    \label{fig:performance_eval}
\end{figure}

In Figure \ref{fig:performance_eval}, we show the averaged curves across all the folds considering the performance from four perspectives.




The validation curves demonstrate consistent improvements in all metrics during the initial training epochs (0–5), followed by stabilization around epoch 10, indicating efficient learning and convergence. Specifically, the sharp decrease in loss highlights rapid optimization during early epochs. The MSSIM curves approaching values near 1.0 suggest excellent preservation of structural similarity, essential for retaining image integrity. Similarly, the decreasing Relative L2 Norm indicates that the network output aligns more closely with the noisy input while minimizing residual differences. The stabilization of metrics after epoch 10 suggests that the network has reached an optimal balance between noise suppression and structural preservation. Finally, the steady increase in Proxy PSNR reflects the network's ability to suppress noise effectively, reaching the value of around 40 dB. These trends are driven by the network's iterative framework, which allows for progressive refinement through contextual integration and residual learning, enabling it to suppress noise while retaining critical image details effectively. This unsupervised learning approach enhances the practical applicability of CPSNet in reconstructing LUS images, where acquiring clean, annotated data is often challenging.

\subsection{B-line Identification}
\label{sec:identification}
For B-line identification, we report the performance of the 10-layer CPSNet as a representative evaluation of the model's effectiveness. Table \ref{tab:lineDec} summarizes the quantitative results. 

CPSNet demonstrates superior performance in B-line identification, achieving the highest F1 score (0.608 ± 0.023) and F2 score (0.621 ± 0.047), along with the second best recall (62.93\% ± 6.67\%), showcasing its ability to detect B-lines accurately and consistently. When compared to the baseline method PUI, CPSNet achieves significant improvements, with an 11.93\% increase in precision and 17.56\% improvement in recall, demonstrating better sensitivity while maintaining competitive precision. Against the traditional CPS method, CPSNet achieves a 0.141 improvement in F1 score and a substantial recall boost, highlighting its enhanced robustness. 

Compared to other neural networks such as FasterRCNN and YOLO, CPSNet not only achieves superior performance but also exhibits far greater consistency. While Faster R-CNN reports the highest mean precision (78.37 \% ± 23.47 \%), this figure is misleading in the context of clinical diagnostics. In the 3-fold cross-validation (repeated twice), some folds contained almost no predictions from Faster R-CNN, and in others only a small number of predictions were made—many of which happened to be correct. In these situations, the precision metric becomes artificially inflated, as it only accounts for the ratio of true positives to the sum of true and false positives. This inflation can occur even when a large number of detections are missed. This is reflected in the very low mean recall (26.57 \% ± 31.39 \%) and high variance, indicating that the method fails to detect B-lines in a substantial proportion of cases. Similarly, YOLOv11 is problematic but showing an opposite result. The high recall (79.72\% ± 35.13 \%) comes from an excessive number of predictions, but zero predictions in other folds contributes to the low F1 score. In clinical practice, such behaviour is unacceptable, as missing a large fraction of true pathological signs undermines the diagnostic utility of the system, regardless of its apparent precision in favourable cases. In contrast, CPSNet achieves the best balance across all metrics, with both precision and recall above 58 \% and 62 \% respectively, yielding the highest F1 and F2 scores. Moreover, its narrow standard deviations highlight consistent performance across all folds, making it more reliable for real-world deployment.

In addition to the fixed-threshold metrics, the proxy mAP provides an estimate of performance stability across different score thresholds. CPSNet achieves the highest proxy mAP (0.25 $\pm$ 0.02), outperforming CPS (0.21 $\pm$ 0.02), DUBLINE (0.18 $\pm$ 0.02), and PUI (0.15 $\pm$ 0.01). This suggests that CPSNet maintains a more favourable precision--recall trade-off across the threshold range, rather than being optimised only for a single operating point. In contrast, the proxy mAP values of YOLOv7, YOLOv11, and Faster R-CNN remain low, indicating unstable detection behaviour caused by either excessive false positives or missed detections across folds.

In terms of computational efficiency, CPSNet achieves the fastest execution speed (0.01 seconds/frame), making it the most efficient method in the comparison. It significantly outperforms both traditional approaches and modern deep neural networks. Compared to PUI and CPS, CPSNet eliminates the need for repeated computation of the Radon transform, reducing computational overhead. Additionally, unlike existing deep learning models, CPSNet benefits from a streamlined network architecture with shallower layers, significantly lowering computational load and resource requirements. This efficiency makes CPSNet a practical solution for large-scale applications where high processing speed is essential, further highlighting its effectiveness in B-line detection tasks.

\begin{table}[t!]
\centering
\caption{B-line Identification Performance (Threshold = 0.5). The best scores are in bold, and the second-best scores are underlined.}
\label{tab:lineDec}{%
\resizebox{\textwidth}{!}{
\begin{tabular}{ccccccccc}
\hline
                                   & Precision (\%) $\uparrow$ & Recall (\%) $\uparrow$ & F1 $\uparrow$& F2 $\uparrow$ & \multicolumn{2}{c}{\begin{tabular}[c]{@{}c@{}}Execution Speed $\downarrow$\\ (seconds/frame)\end{tabular}} & proxy mAP $\uparrow$\\ \hline
PUI \cite{anantrasirichai2017line} & 46.97±8.84          & 45.37±2.63       & 0.462±0.037  & 0.457±0.018  & \multicolumn{2}{c}{4.14}   & 0.15±0.01                                                                       \\
CPS \cite{karakucs2020detection} $^{*}$   & 42.86±7.21          & 51.22±4.39       & 0.467±0.044  & 0.493±0.035 & \multicolumn{2}{c}{1.32}   & \underline{0.21±0.02 }                                                                      \\
DUBLINE \cite{yang2023dubline}     & 31.68±5.62          & 61.61±6.93       & 0.413±0.046  & \underline{0.518±0.043} & \multicolumn{2}{c}{0.11}               & 0.18±0.02                                                                \\
YOLOv7 \cite{wang2023yolov7}       & 27.44±20.80          & 31.25±23.78       & 0.324±0.165  & 0.304±0.175 & \multicolumn{2}{c}{0.04}  & 0.03±0.03                                                                               \\
YOLOv11m \cite{yolo11_ultralytics}       & 26.29±37.32          & \textbf{79.72±35.13}       & 0.146±0.098  & 0.205±0.035 & \multicolumn{2}{c}{\underline{0.02}}             & 0.02±0.01                                                                    \\
FasterRCNN \cite{ren2015faster}    & \textbf{78.37±23.47}          & 26.57±31.39       & \underline{0.550±0.106}  & 0.306±0.169 & \multicolumn{2}{c}{0.09}          & 0.03±0.03                                                                    \\
CPSNet 10-Layer (ours)             & \underline{58.90±3.58 }         & \underline{62.93±6.67 }      & \textbf{0.608±0.023}  & \textbf{0.621±0.047} & \multicolumn{2}{c}{\textbf{0.01}}                     & \textbf{0.25±0.02}                                                            \\ \hline
\multicolumn{6}{l}{$^{*}$ Excluding data that exceeds the maximum iteration count. Examples are shown in Figure \ref{fig:mainfig}.}                                                                                                              
\end{tabular}%
}
}
\end{table}

\begin{figure}[t!]
\centering
\includegraphics[width=0.8\linewidth]{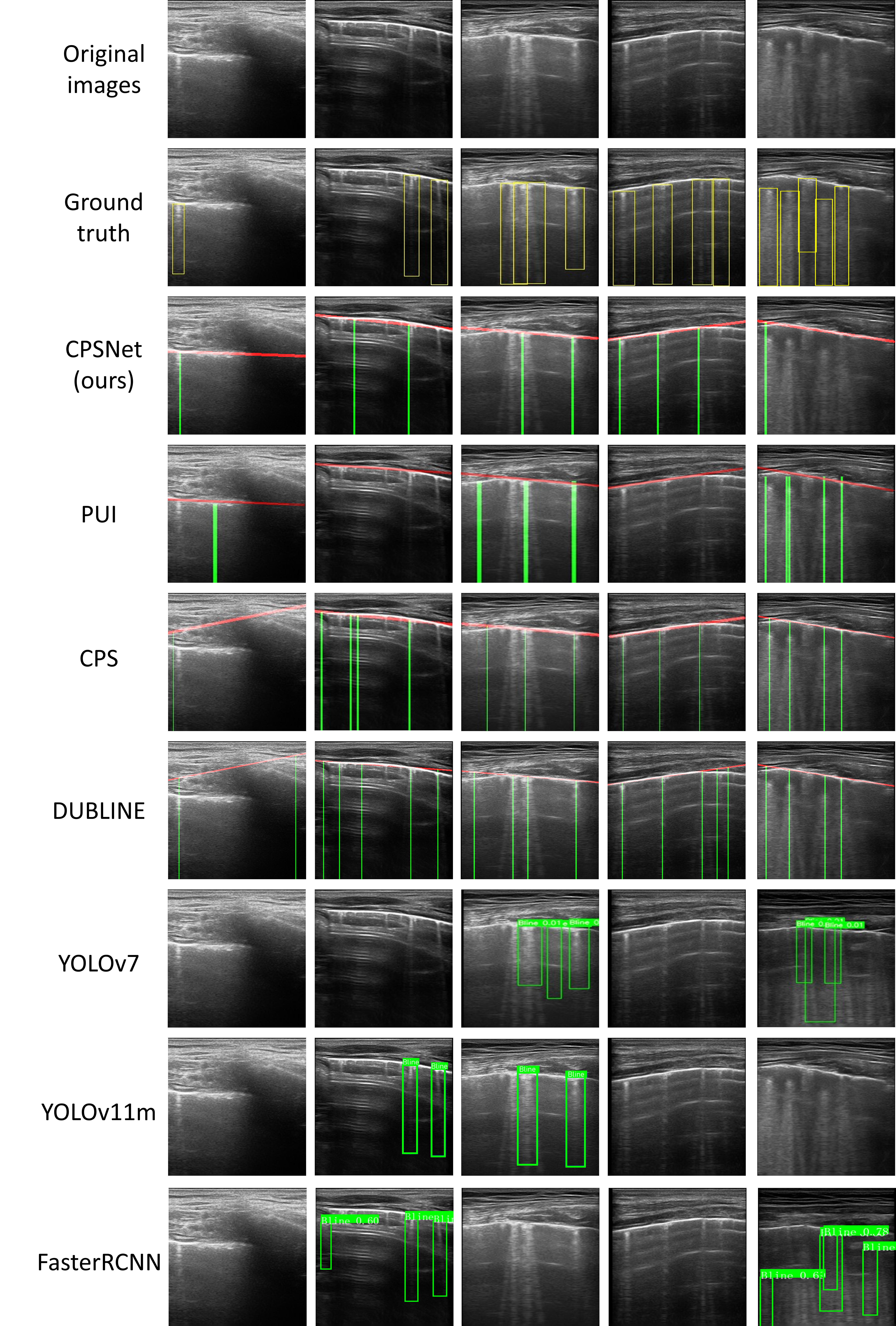}
\caption{Line detection results for various frames from different patients. Ground truth is annotated by yellow boxes. For model-based methods and deep unfolding methods, pleural lines are depicted as red lines, and detected B-lines are shown as green lines. For YOLOv7, YOLOv11m and FasterRCNN, detected B-lines are shown in green boxes.}
\label{fig:compare}
\end{figure}

\begin{figure}[t!]
    \centering
    \begin{subfigure}[t]{0.3\textwidth}
        \centering
        \includegraphics[width=\textwidth]{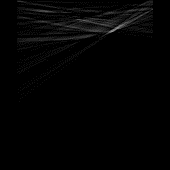}
        \caption{}
        \label{fig:subfig1}
    \end{subfigure}
    \begin{subfigure}[t]{0.3\textwidth}
        \centering
        \includegraphics[width=\textwidth]{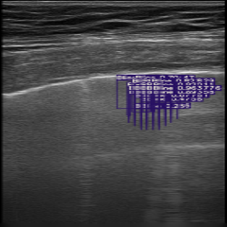}
        \caption{}
        \label{fig:subfig2}
    \end{subfigure}
    \begin{subfigure}[t]{0.3\textwidth}
        \centering
        \includegraphics[width=\textwidth]{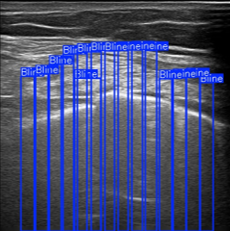}
        \caption{}
        \label{fig:subfig3}
    \end{subfigure}
    
    \caption{Examples of false cases. (a) CPS sometimes over-regularize the, (b) YOLOv7 identifies wrong features as B-lines, and (c) YOLOv11m predicts excessive number of vertical artefacts.}
    \label{fig:mainfig}
\end{figure}

Subjective evaluation is shown in Figure \ref{fig:compare}. CPSNet demonstrates higher accuracy in frames with relatively more sparse B-lines (column 1, 2, 3 and 4), where it captures the essential line structures with minimal false positives (FPs). However, in frames with a high density of B-lines (column 5), CPSNet misses some detections, likely due to the increased complexity of the overlapping structures. This tendency to under-detect in more complex frames may contribute to its relatively low precision score. In contrast, model-based methods like PUI and CPS show relatively consistent detections that align with ground truth especially when there are more B-lines. Object detection models like YOLO and FasterRCNN exhibit variability, with some frames showing accurate detections and others missing B-lines entirely or generating FPs. YOLOv7, for instance, occasionally detects artifacts or irrelevant structures as B-lines, indicating a sensitivity to noise in ultrasound images (see Figure \ref{fig:subfig2}).

A key observation in the evaluation is the over-regularization problem encountered in traditional CPS optimization. Traditional CPS applies a fixed regularization term to suppress noise, but this often leads to excessive smoothing or over-regularization, where true B-lines are diminished and fake information is introduced (as shown in Figure \ref{fig:subfig1}). This results in false detections in 118 out of 360 test images. The problem reduces the diagnostic reliability of the traditional CPS method, as the overly smoothed output may mislead clinicians by highlighting non-existent B-lines or concealing subtle yet clinically relevant ones.

CPSNet overcomes this issue through its adaptive regularization that adjusts to the unique characteristics of each image. Instead of applying a fixed regularization parameter, CPSNet learns an optimal balance between noise suppression and feature preservation from the data itself, allowing it to retain genuine B-lines while avoiding the false detections caused by over-regularization. This adaptability makes CPSNet more robust across diverse images, preserving clinically relevant features and reducing the risk of erroneous B-line detections.

\subsection{Ablation Study}
\label{sec:ablation}

\begin{table}[t!]
\centering
\caption{Impact of Number of Layers on B-line Identification (Threshold=0.5).The best scores are in bold.}
\label{tab:num_layer}
{%
\resizebox{0.9\textwidth}{!}{
\begin{tabular}{ccccc}
\hline
Number of Layer & Precision (\%) $\uparrow$ & Recall (\%) $\uparrow$ & F1 $\uparrow$ & F2 $\uparrow$\\ \hline
1         & 57.56±3.26          & 62.20±7.41       & 0.598±0.033  & 0.612±0.055 \\
2         & 57.69±3.39          & 62.20±7.78       & 0.599±0.035  & 0.612±0.058 \\
3         & 58.37±3.73          & 62.93±7.07       & 0.606±0.030  & 0.620±0.051 \\
4         & 58.14±3.35          & 62.68±6.90       & 0.603±0.027  & 0.617±0.049 \\
5         & 58.41±3.51          & 62.68±6.90       & 0.605±0.027  & 0.618±0.049 \\
6         & 57.69±3.19          & 62.20±7.41       & 0.599±0.032  & 0.612±0.055 \\
7         & 57.60±3.03          & 61.95±6.62       & 0.597±0.019  & 0.610±0.045 \\
8         & 58.41±3.08          & 62.68±7.36       & 0.605±0.028  & 0.618±0.053 \\
9         & 58.05±3.10          & 62.44±7.14       & 0.602±0.029  & 0.615±0.052 \\
10        & \textbf{58.90±3.58}          & \textbf{62.93±6.67}       & \textbf{0.608±0.023}  & \textbf{0.621±0.047} \\ \hline
\end{tabular}%
}
}
\end{table}

\begin{table}[t!]
\centering
\caption{Execution Speed for Different Number of Layers in CPSNet (seconds/frame)}
\label{tab:time}
\resizebox{\textwidth}{!}{%
\begin{tabular}{cccccccccc|c}
\hline
1-Layer & 2-Layer & 3-Layer & 4-Layer & 5-Layer & 6-Layer & 7-Layer & 8-Layer & 9-Layer & 10-Layer & Average \\\hline
0.0213  & 0.0202  & 0.0126  & 0.0111  & 0.0118  & 0.0119  & 0.0124  & 0.0122  & 0.0119  & 0.0112   & 0.0136  \\ \hline
\end{tabular}%
}
\end{table}

\begin{table}[t!]
\centering
\caption{Impact of the Cauchy Regularization on B-line Identification (10-Layer, Threshold=0.5)}
\label{tab:losscompare}
\resizebox{0.8\textwidth}{!}{%
\begin{tabular}{ccccc}
\hline
               & Precision (\%) $\uparrow$   & Recall (\%) $\uparrow$     & F1  $\uparrow$   & F2  $\uparrow$      \\  \hline
$\lambda = 0.01$ & 58.90±3.58 & 62.93±6.67 & 0.608±0.023 & 0.621±0.047 \\
$\lambda = 0.00$ & 58.41±3.14 & 62.68±6.90 & 0.605±0.026 & 0.618±0.049\\ \hline
\end{tabular}
}
\end{table}

\textbf{Impact of Numbers of Layers}: In Figure \ref{fig:performance_eval}, rather than displaying results for all possible layer configurations, we present representative curves corresponding to network structures with 1, 5, and 10 layers for the clarity of comparison. This selection provides a concise visualization while preserving the overall performance trends. On analysing the plots, it can be seen that architectures with different numbers of layers show variations in how quickly the metrics improve at the beginning. However, as training progresses, structural fidelity and noise suppression improve across all layer configurations but plateaus over time. Finally, the metrics all converge to similar levels regardless of the number of layers. This indicates that while additional layers may impact the early phases of optimization, it has a minimal effect on the network's final performance. 

Table \ref{tab:num_layer} reports the impact of the number of layers on the performance of CPSNet for B-line identification. The results show that while increasing the number of layers leads to slight improvements in metrics, the overall impact remains minimal. For instance, the F1 score improves marginally from 0.598 ± 0.033 with 1 layer to 0.608 ± 0.023 with 10 layers, and the Recall increases from 62.20\% ± 7.41\% to 62.93\% ± 6.67\%. These incremental changes suggest that CPSNet achieves strong performance even with fewer layers, reflecting the robustness and efficiency of its architecture.

The limited influence of layer count can be attributed to the iterative design of CPSNet, where each layer refines the solution in a structured manner. Even a single layer effectively captures the essential features for B-line identification. This reveals that CPSNet does not rely heavily on depth for performance but rather on its efficient integration of forward and backward blocks, which balance noise suppression and structural preservation. Furthermore, the inclusion of skip connections ensures the propagation of critical information, reducing reliance on additional layers and avoiding unnecessary complexity that offers minimal performance gains.

Table \ref{tab:time} demonstrates that CPSNet achieves efficient execution speeds in various layer configurations, with processing times stabilizing around 0.012 seconds/frame for networks with 4 or more layers. The 1-layer configuration takes the longest (0.0213 seconds/frame), likely due to less efficient feature refinement requiring more iterative computations. As the number of layers increases, the network leverages its modular architecture and optimized iterative framework to perform feature extraction and refinement more efficiently, resulting in faster execution. The stabilization of execution speed beyond 4 layers suggests that the additional layers contribute marginally to computational overhead, possibly due to the effectiveness of skip connections in propagating essential information without redundancy. These results reflect the careful balance between network depth and processing efficiency, making CPSNet both practical and reliable for tasks requiring high-speed performance in resource-constrained environments.

While deeper networks can provide more refinement steps, two practical considerations motivated the decision not to extend the depth beyond 10 layers:
\begin{enumerate}
\item Balancing computational efficiency and performance – Although inference time remains effectively constant due to parallel execution on the GPU, deeper unrolled networks require proportionally more training time and memory during optimisation. Given that the observed performance gain from 6 to 10 layers is small (F1-score improvement of only ~0.01), the added training cost was not justified for this application.

\item Consistency with traditional iterative algorithms – In the classical forward–backward scheme, iterations are terminated when a stopping criterion is met (e.g., the update error falls below a small threshold) or when a maximum number of iterations is reached to avoid unnecessary computation if convergence stalls. Limiting CPSNet to 10 unfolded layers mirrors this practice: it provides sufficient refinement steps to reach a stable solution without excessive iteration, while preventing overfitting to noise in the unsupervised training setting.
\end{enumerate}
We therefore regard 10 layers as the optimal depth used in this study, as it achieves a balance between performance, training efficiency, and adherence to the theoretical behaviour of the underlying iterative algorithm.

\textbf{Impact of Cauchy Regularization}: 
Table \ref{tab:losscompare} evaluates the impact of Cauchy regularization on the performance of CPSNet for B-line identification. The results compare the network's performance with and without the Cauchy regularization term. The inclusion of Cauchy regularization leads to slight improvements across all performance metrics in terms of mean values and appears to reduce performance variability. These findings suggest that the Cauchy penalty enhances both the accuracy and stability of the network.

The improvement can be attributed to the properties of Cauchy regularization, which introduces a robust sparsity-promoting mechanism in the network. By penalizing large deviations while being less sensitive to small ones, the Cauchy regularization helps the model refine the optimization process by discouraging over-reliance on high-intensity features or noise in the Radon domain. Without regularization, the network relies solely on the reconstruction loss terms, potentially leading to less robust performance in noisy scenarios. However, the relatively small impact of the regularization can be explained by the already robust architecture of CPSNet and the choice of the method in the B-line identification stage.

\section{Conclusion}
\label{sec:conclusion}
In this work, we introduced CPSNet, a novel framework designed for robust and efficient LUS image reconstruction and B-line detection, balancing speed, stability, and adaptability. Leveraging a deep unfolding structure, CPSNet iteratively enhances images by reducing noise while preserving critical structures, avoiding the over-regularization often observed in traditional optimization methods. This adaptability allows CPSNet to retain essential B-line features, a crucial advantage for clinical applications. Additionally, its architectural design enables effective feature refinement without requiring excessive depth, ensuring computational efficiency without compromising accuracy. Moreover, CPSNet exhibits low variability across cross-validation folds, demonstrating consistent performance superior to the high variability observed in object detection models such as YOLO and FasterRCNN. Its rapid execution time further reinforces its suitability for real-time applications, making it an efficient and reliable tool for point-of-care diagnostics.\\

\section{Ethical Statement}
The study protocol conformed to the Declaration of Helsinki and was approved by a local research ethics committee (study approval number 18217/OSS). Informed consent was obtained from all subjects involved in the study.

\bibliographystyle{elsarticle-num} 
\bibliography{ref}
\end{document}